# Projected Iterative Soft-thresholding Algorithm for Tight Frames in Compressed Sensing Magnetic Resonance Imaging

Yunsong Liu, Zhifang Zhan, Jian-Feng Cai, Di Guo, Zhong Chen, Xiaobo Qu*

*Abstract*—Compressed sensing has shown great potentials in accelerating magnetic resonance imaging. Fast image reconstruction and high image quality are two main issues faced by this new technology. It has been shown that, redundant image representations, e.g. tight frames, can significantly improve the image quality. But how to efficiently solve the reconstruction problem with these redundant representation systems is still challenging. This paper attempts to address the problem of applying iterative soft-thresholding algorithm (ISTA) to tight frames based magnetic resonance image reconstruction. By introducing the canonical dual frame to construct the orthogonal projection operator on the range of the analysis sparsity operator, we propose a projected iterative soft-thresholding algorithm (pISTA) and further accelerate it by incorporating the strategy proposed by Beck and Teboulle in 2009. We theoretically prove that pISTA converges to the minimum of a function with a balanced tight frame sparsity. Experimental results demonstrate that the proposed algorithm achieves better reconstruction than the widely used synthesis sparse model and the accelerated pISTA converges faster or comparable to the state-of-art smoothing FISTA. One major advantage of pISTA is that only one extra parameter, the step size, is introduced and the numerical solution is stable to it in terms of image reconstruction errors, thus allowing easily setting in many fast magnetic resonance imaging applications.

This work was supported by the NNSF of China (61571380, 61201045, 61302174 and 11375147), Natural Science Foundation of Fujian Province of China (2015J01346), Fundamental Research Funds for the Central Universities (20720150109, 2013SH002) and NSF DMS-1418737. (*Corresponding author: Xiaobo Qu)

Yunsong Liu, Zhifang Zhan, Zhong Chen and Xiaobo Qu are with the Department of Electronic Science, Fujian Provincial Key Laboratory of Plasma and Magnetic Resonance, Xiamen University, Xiamen, China (e-mail: yunsongliu@stu.xmu.edu.cn; zhanzfadu@stu.xmu.edu.cn; chenz@xmu.edu.cn; quxiaobo@xmu.edu.cn).
Jian-Feng Cai is with Department of Mathematics, Hong Kong University of Science and Technology, Hong Kong SAR, China. This work was partially done when J. F. Cai was at Department of Mathematics, University of Iowa, Iowa City, USA. (Email: jfcai@ust.hk)
Di Guo is with the School of Computer and Information Engineering, Fujian Provincial University Key Laboratory of Internet of Things Application Technology, Xiamen University of Technology, Xiamen, China (e-mail: guodi@xmut.edu.cn)

## I. Introduction

MAGNETIC resonance imaging (MRI) plays an important role in clinical diagnosis nowadays. MRI is noninvasive and can provide high quality images, especially for soft tissues. One major challenge that MRI faces is the fundamental limitation of its imaging speed [1] both physically (e.g. gradient amplitude and slew-rate), and physiologically (e.g. nerve stimulation). Compressed sensing (CS) [1-3] is a promising technique to accelerate MRI by undersampling k-space data, better known as Fourier space data. This new technology is called compressed sensing MRI [1], or CS-MRI for short. The undersampling process can be mathematically modeled as

$$\mathbf{y} = \mathbf{UFx} + \mathbf{\eta} , \qquad (1)$$

where $\mathbf{x} \in \mathbb{C}^N$ represents the magnetic resonance (MR) image rearranged to a column vector, $\mathbf{F} \in \mathbb{C}^{N \times N}$ is the discrete Fourier transform, $\mathbf{U} \in \mathbb{R}^{M \times N} (M < N)$ is the undersampling matrix, $\mathbf{\eta} \in \mathbb{C}^M$ is the additive noise, and $\mathbf{y} \in \mathbb{C}^M$ is the undersampled k-space data. The goal of image reconstruction is to recover a reasonable $\mathbf{x}$ with $N$ pixels from $\mathbf{y}$ with $M$ data points, which is ill-posed. CS-MRI solves this problem by assuming that MR images have a good sparse approximation under a certain transform such as finite difference or a wavelet transform[1]. Mathematically, sparsity is promoted directly by minimizing the $\ell_0$ norm of coefficients in the transform domain. However, minimization involving the lo-norm is generally NP-hard and thus intractable[3]. The $\ell_1$ norm is a good relaxation that can work as well as $\ell_0$ norm under some conditions, and more practically the reconstruction problems can be solved efficiently with convex optimization methods [2, 3].

One key to the success of CS-MRI is the design of systems that can sparsify MR images. Some well-established such systems include traditional wavelets [1], contourlets [4], directional wavelets [5] and emerging trained dictionaries [6], etc. The latter image representation systems have shown to improve the MR image reconstruction on edges and artifacts removal. In the signal processing, these systems can be well defined with the concept of tight frames [7]. To allow rigorously and clearly define the MR image reconstruction problems to be solved in this paper, we first briefly review the definition of tight frames; see [7] for more details.

A set of vectors $\{\mathbf{d}_j\}_{j=1,2,\ldots,J}$ is called a frame in $\mathbb{C}^N$ if there exist positive real numbers $A, B$ such that [7]

$$A\|\mathbf{x}\|_2^2 \leq \sum_j \left|\langle \mathbf{x}, \boldsymbol{\varphi}_j \rangle\right|^2 \leq B\|\mathbf{x}\|_2^2 \quad \text{for every } \mathbf{x} \in \mathbb{C}^N, \quad (2)$$

where $\langle \mathbf{x}, \mathbf{d}_j \rangle = \mathbf{d}_j^* \mathbf{x}$ is the inner product between $\mathbf{x}$ and $\mathbf{d}_j$, and $\mathbf{d}_j^*$ represents the Hermitian transpose of $\mathbf{d}_j$. A frame is complete to represent any signals lie in N-dimensional Hilbert space while the uniqueness of the representation is not necessary.

When $A = B$, we call this frame is a tight frame. For example, both the contourlets [4] and directional wavelets [5] are tight frames.

The synthesis and analysis operators associated with the frame $\{\mathbf{d}_j\}_{j=1,2,\ldots,J}$ are defined as

$$\begin{aligned} \text{Synthesis operator:} \quad & \mathbf{D} = [\mathbf{d}_1, \mathbf{d}_2, \ldots, \mathbf{d}_J] \\ \text{Analysis operator:} \quad & \boldsymbol{\Psi} = \mathbf{D}^* \end{aligned} \quad (3)$$

In the rest of this paper, we will also use the synthesis operator $\mathbf{D}$ to denote a frame. Another frame $\boldsymbol{\Phi} = \{\boldsymbol{\varphi}_j\}_{j=1,2,\ldots,J}$ is called a dual frame of $\mathbf{D}$ if $\mathbf{D}\boldsymbol{\Phi}^* = \mathbf{I}$. For a given frame, although there are many dual frames, there is an easy choice called the canonical dual frame defined as

$$\boldsymbol{\Phi} = \left(\boldsymbol{\Psi}^*\boldsymbol{\Psi}\right)^{-1}\boldsymbol{\Psi}^*, \quad (4)$$

which is also known as the pseudo-inverse of $\boldsymbol{\Psi}$ [8], satisfying
$$\boldsymbol{\Phi}\boldsymbol{\Psi} = \mathbf{I}. \quad (5)$$
An important property of the canonical dual frame is that we can easily construct the orthogonal projection on $\text{Range}(\boldsymbol{\Psi}) = \{\boldsymbol{\Psi}\mathbf{x} \mid \mathbf{x} \in \mathbb{C}^N\}$ according to

$$P = \boldsymbol{\Psi}\boldsymbol{\Phi} \quad (6)$$

as illustrated in Fig. 1.

The projection operator $P$ has the property that
$$P\boldsymbol{\alpha} = \boldsymbol{\Psi}\boldsymbol{\Phi}\boldsymbol{\alpha} = \boldsymbol{\alpha}, \quad (7)$$
for $\boldsymbol{\alpha} \in \text{Range}(\boldsymbol{\Psi})$.

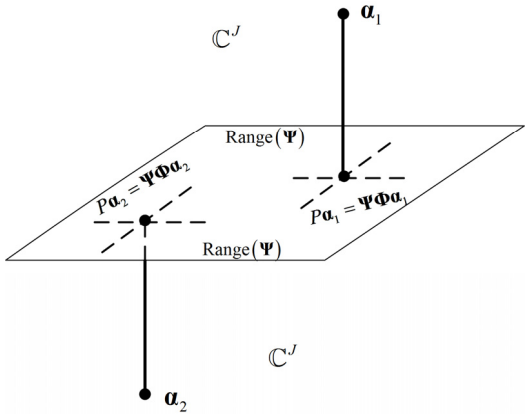

Fig. 1. Illustration of the orthogonal projection operation on $\text{Range}(\boldsymbol{\Psi})$.

In this paper, we focus on tight frames with constant $A = B = 1$. It is easy to see that any tight frame can be rescaled to $A = B = 1$ without affecting its property. The dual frame of such a tight frame is itself, namely $\boldsymbol{\Psi}^*\boldsymbol{\Psi} = \mathbf{I}$.

When we use $\ell_1$ norm to promote sparsity under tight frame representations, there are two different models, namely, the synthesis model and the analysis model [8-10]:

$$\text{Synthesis:} \quad \min_{\boldsymbol{\alpha}} \lambda \|\boldsymbol{\alpha}\|_1 + \frac{1}{2}\|\mathbf{y} - \mathbf{UF}\boldsymbol{\Phi}\boldsymbol{\alpha}\|_2^2 \quad (8)$$

$$\text{Analysis:} \quad \min_{\mathbf{x}} \lambda \|\boldsymbol{\Psi}\mathbf{x}\|_1 + \frac{1}{2}\|\mathbf{y} - \mathbf{UF}\mathbf{x}\|_2^2, \quad (9)$$

where $\boldsymbol{\alpha}$ is the coefficient that is synthesized by $\boldsymbol{\Phi}$ to be an image, meaning that $\mathbf{x} = \boldsymbol{\Phi}\boldsymbol{\alpha}$. Here $\lambda$ is called the regularization parameter which balances a trade-off between sparsity and data fidelity. When $\boldsymbol{\Psi}$ is an orthonormal basis and $\boldsymbol{\Phi} = \boldsymbol{\Psi}^* = \boldsymbol{\Psi}^{-1}$ ($\boldsymbol{\Psi}^*$ and $\boldsymbol{\Psi}^{-1}$ denote the Hermitian transpose and the inverse of $\boldsymbol{\Psi}$, respectively), the analysis and synthesis models yield same solutions [8]. Compared to orthogonal systems, redundant systems, such as tight frames and dictionaries, can benefit from that redundancy in noise removal and artifacts reduction in signal processing [10-12]. Within the field of CS-MRI, the quality of reconstructed images is improved with redundant systems [6, 13-16]. For these redundant systems, there exists significant difference between these two models as reported in [8, 14, 17, 18]. Theoretically, CS theory [10] reveals that analysis models works under a wider range of dictionaries than the synthesis models, and it was proved in [19] that analysis models converge to some partial differential equation models with geometric interpretations. In practice, researchers in signal processing observed that analysis models reaches lower reconstructed error than synthesis models [8, 14, 17, 18].

Another challenge in CS-MRI is to design fast and efficient algorithms to solve the convex optimization problems especially with redundant sparse representation systems. Iterative soft-thresholding algorithm (ISTA), also known as proximal forward backward splitting algorithm, is a kind of simple and efficient algorithm for the $\ell_1$-norm based sparse recovery problems [20-25]. Beck and Teboulle [26] proposed a fast ISTA (FISTA) by utilizing a very specific linear combination of the previous two iterations to significantly speed up the convergence of ISTA [26]. The efficiencies of both ISTA and FISTA depend on the simplicity of computing the proximal map of the non-smooth $\ell_1$ norm based functions in objectives. The proximal map of a function $f$ is defined as [22, 27]

$$\text{prox}_f(\mathbf{x}) = \arg\min_{\mathbf{z}} f(\mathbf{z}) + \frac{1}{2}\|\mathbf{z} - \mathbf{x}\|_2^2. \quad (10)$$

For a synthesis model in (8), $f(\boldsymbol{\alpha}) = \lambda\|\boldsymbol{\alpha}\|_1$, the proximal map is

$$\text{prox}_f(\boldsymbol{\alpha}) = \arg\min_{\mathbf{z}} \lambda\|\mathbf{z}\|_1 + \frac{1}{2}\|\mathbf{z} - \boldsymbol{\alpha}\|_2^2 = T_\lambda(\boldsymbol{\alpha}), \quad (11)$$

where $T_\lambda(\cdot)$ is a point wise soft-thresholding function as

$$T_\lambda(\alpha_i) = \max\{|\alpha_i| - \lambda, 0\} \cdot \frac{\alpha_i}{|\alpha_i|}. \quad (12)$$

The simplicity of the proximal map (11) makes FISTA very efficient for solving synthesis models. Whereas for analysis models, there is no such simple closed form solution for the proximal map of $f(\mathbf{x}) = \lambda \|\mathbf{\Psi x}\|_1$, resulting in challenges to apply ISTA and FISTA to analysis models.

To address this problem, Tan., et al. [28] proposed smoothing FISTA (SFITSA) for tight frames in sparse image reconstruction [28]. The convergence was analyzed and a performance bound on the reconstruction error was derived there. SFISTA replaces the non-smooth term $f(\mathbf{x}) = \lambda \|\mathbf{\Psi x}\|_1$ by its Moreau envelop, defined as [28]

$$f_\mu(\mathbf{x}) = \min_{\mathbf{z}} \lambda \|\mathbf{\Psi z}\|_1 + \frac{1}{2\mu} \|\mathbf{z} - \mathbf{x}\|_2^2, \quad (13)$$

which is smooth. Then, they applied FISTA to solve this relaxed smooth optimization problem. As it is shown in [28] and our numerical experiments, both the convergence speed and reconstructed error of SFISTA are sensitive to the smooth approximate parameter $\mu$. By introducing the continuation strategy to gradually decreasing the value of $\mu$, Tan., et al. argue that the solution of the previous problem with larger $\mu$ will provide a warm start of the current problem with smaller $\mu$ and thus can make SFISTA converge fast with high accuracy of the smooth approximation [28]. Despite the effectiveness of this continuation strategy as demonstrated in [28], it introduces more parameters, such as the decrease rate, the maximum number of iterations for a sub-problem with a specific value of $\mu$. This makes SFISTA relatively complicated for CS-MRI applications.

In this paper, we apply a variant of ISTA and FISTA to approximately solve the analysis model for tight frames in CS-MRI. With the canonical dual frame, we rewrite the analysis model to be a constrained synthesis-like one. This inspires us to apply algorithms that are usually fit for synthesis models, e.g. ISTA, to analysis models. In order to keep the simplicity of ISTA, we propose to replace a constrained proximal map by an unconstrained proximal map plus the orthogonal projection in (6). Therefore, the proposed algorithm is called projected ISTA (pISTA). Furthermore, the same accelerating strategy as FISTA [26] is introduced, resulting in the projected FISTA (pFISTA). We analyze the convergence of pISTA and pFISTA in section II(B). Then, in section III we verify the efficiency of pFISTA by comparing it with FISTA, solving the synthesis model, and the state-of-the-art SFISTA, solving the analysis model.

## II. Proposed Method

### A. Projected Iterative Soft-Thresholding Algorithm

The analysis model in (9) is equivalent to the following form

$$\min_{\boldsymbol{\alpha} \in \text{Range}(\mathbf{\Psi})} \lambda \|\boldsymbol{\alpha}\|_1 + \frac{1}{2} \|\mathbf{y} - \mathbf{UF\Phi\alpha}\|_2^2. \quad (14)$$

This equivalence means that the solutions of (9) and (14) are the same. The proof of this equivalence is in the Appendix.

To handle the constraint in the synthesis-like analysis model in (14), we introduce an indicator function

$$d(\boldsymbol{\alpha}) = \begin{cases} 0 & , \boldsymbol{\alpha} \in \text{Range}(\mathbf{\Psi}) \\ \infty & , \boldsymbol{\alpha} \notin \text{Range}(\mathbf{\Psi}) \end{cases} \quad (15)$$

to obtain an equivalent unconstrained model of (14) as

$$\min_{\boldsymbol{\alpha}} \lambda \|\boldsymbol{\alpha}\|_1 + \frac{1}{2} \|\mathbf{y} - \mathbf{UF\Phi\alpha}\|_2^2 + d(\boldsymbol{\alpha}). \quad (16)$$

We further denote that

$$\begin{aligned} g(\boldsymbol{\alpha}) &= \lambda \|\boldsymbol{\alpha}\|_1 + d(\boldsymbol{\alpha}) \\ f(\boldsymbol{\alpha}) &= \frac{1}{2} \|\mathbf{y} - \mathbf{UF\Phi\alpha}\|_2^2 \end{aligned} \quad (17)$$

where $g(\boldsymbol{\alpha})$ is a non-smooth convex function, and $f(\boldsymbol{\alpha})$ is a smooth function with a $L_f$-Lipschitz continuous gradient $\nabla f$, i.e.

$$\|\nabla f(\boldsymbol{\alpha}_1) - \nabla f(\boldsymbol{\alpha}_2)\|_2 \leq L_f \|\boldsymbol{\alpha}_1 - \boldsymbol{\alpha}_2\|_2 \quad (18)$$

where $L_f > 0$ [27]. Then, we apply proximal algorithm [19, 24] to solve the problem in (16) by incorporating the proximal mapping

$$\begin{aligned} \boldsymbol{\alpha}_{k+1} &= \text{prox}_{\gamma g}(\boldsymbol{\alpha}_k - \gamma \nabla f(\boldsymbol{\alpha}_k)) \\ &= \arg\min_{\boldsymbol{\alpha}} \lambda\gamma \|\boldsymbol{\alpha}\|_1 + \frac{1}{2} \|\boldsymbol{\alpha} - (\boldsymbol{\alpha}_k - \gamma \nabla f(\boldsymbol{\alpha}_k))\|_2^2 + d(\boldsymbol{\alpha}), \quad (19) \\ &= \arg\min_{\boldsymbol{\alpha} \in \text{Range}(\mathbf{\Psi})} \lambda\gamma \|\boldsymbol{\alpha}\|_1 + \frac{1}{2} \|\boldsymbol{\alpha} - (\boldsymbol{\alpha}_k - \gamma \nabla f(\boldsymbol{\alpha}_k))\|_2^2 \end{aligned}$$

where $\gamma$ is the step size.

So far, we have converted the original analysis model based CS-MRI problem into a much simpler form in (19) where the objective function is separable. However, the constraint $\boldsymbol{\alpha} \in \text{Range}(\mathbf{\Psi})$ makes us hard to find an analytical solution of (19). Observing that without this constraint, (19) degenerates to the proximal mapping in (11) and its closed form solution is

$$\tilde{\boldsymbol{\alpha}}_{k+1} = T_{\gamma\lambda}\left((\boldsymbol{\alpha}_k - \gamma \nabla f(\boldsymbol{\alpha}_k))\right). \quad (20)$$

We propose to replace (19) by

$$\begin{aligned} \tilde{\boldsymbol{\alpha}}_{k+1} &= T_{\gamma\lambda}\left((\boldsymbol{\alpha}_k - \gamma \nabla f(\boldsymbol{\alpha}_k))\right) \\ \boldsymbol{\alpha}_{k+1} &= P_{\text{Range}(\mathbf{\Psi})}(\tilde{\boldsymbol{\alpha}}_{k+1}) \end{aligned} \quad (21)$$

where $P_{\text{Range}(\mathbf{\Psi})}$ is the orthogonal projection operator on the $\text{Range}(\mathbf{\Psi})$. More specifically, for our problem in (14), this replacement leads to

$$\begin{aligned} \tilde{\boldsymbol{\alpha}}_{k+1} &= T_{\gamma\lambda}\left((\boldsymbol{\alpha}_k + \gamma \mathbf{\Psi}(\mathbf{\Psi}^*\mathbf{\Psi})^{-1}\mathbf{F}^*\mathbf{U}^T(\mathbf{y} - \mathbf{UF\Phi\alpha}_k))\right) \\ \boldsymbol{\alpha}_{k+1} &= \mathbf{\Psi\Phi}\tilde{\boldsymbol{\alpha}}_{k+1} \end{aligned} \quad (22)$$

The two steps in (22) can be recast as

$$\tilde{\boldsymbol{\alpha}}_{k+1} = T_{\gamma\lambda}\left(\mathbf{\Psi}\left(\mathbf{\Phi}\tilde{\boldsymbol{\alpha}}_k + \gamma(\mathbf{\Psi}^*\mathbf{\Psi})^{-1}\mathbf{F}^*\mathbf{U}^T(\mathbf{y} - \mathbf{UF\Phi}\tilde{\boldsymbol{\alpha}}_k)\right)\right) \quad (23)$$

Furthermore, by substituting the coefficients $\tilde{\boldsymbol{\alpha}}_k$ and $\tilde{\boldsymbol{\alpha}}_{k+1}$ with images $\mathbf{x}_k = \mathbf{\Phi}\tilde{\mathbf{\alpha}}_k$ and $\mathbf{x}_{k+1} = \mathbf{\Phi}\tilde{\mathbf{\alpha}}_{k+1}$, we get that

$$\mathbf{x}_{k+1} = \mathbf{\Phi}T_{\gamma\lambda}\left(\mathbf{\Psi}\left(\mathbf{x}_k + \gamma\left(\mathbf{\Psi}^*\mathbf{\Psi}\right)^{-1}\mathbf{F}^*\mathbf{U}^T\left(\mathbf{y} - \mathbf{UFx}_k\right)\right)\right). \quad (24)$$

For a tight frame, we have $\mathbf{\Phi} = \mathbf{\Psi}^*$ and $\mathbf{\Psi}^*\mathbf{\Psi} = \mathbf{I}$, then (24) becomes

$$\mathbf{x}_{k+1} = \mathbf{\Psi}^*T_{\gamma\lambda}\left(\mathbf{\Psi}\left(\mathbf{x}_k + \gamma\mathbf{F}^*\mathbf{U}^T\left(\mathbf{y} - \mathbf{UFx}_k\right)\right)\right). \quad (25)$$

All the above derivations lead to the proposed projected iterative soft-thresholding algorithm (pISTA) for tight frames based CS-MRI problems. Furthermore, the same accelerating strategy as FISTA [26] is introduced resulting in the projected FISTA (pFISTA). Both pISTA and pFISTA for tight frames in CS-MRI are summarized in **Algorithm 1** and **Algorithm 2**.

For comparison purpose, we list the core iterations of FISTA [26], SFISTA [28] and the proposed pFISTA as follows:

FISTA: $\quad\mathbf{\alpha}_{k+1} = T_{\gamma\lambda}\left(\left(\mathbf{\alpha}_k + \gamma\mathbf{\Psi F}^*\mathbf{U}^T\left(\mathbf{y} - \mathbf{UF\Psi}^*\mathbf{\alpha}_k\right)\right)\right)$

SFISTA: $\quad\mathbf{x}_{k+1} = \left(1 - \gamma/\mu\right)\mathbf{x}_k - \left(\gamma/\mu\right)\mathbf{\Psi}^*T_{\lambda\mu}\left(\mathbf{\Psi x}_k\right)$
$\quad\quad\quad\quad + \gamma\mathbf{F}^*\mathbf{U}^T\left(\mathbf{y} - \mathbf{UFx}_k\right)$ $\quad(26)$

pFISTA: $\quad\mathbf{x}_{k+1} = \mathbf{\Psi}^*T_{\gamma\lambda}\left(\mathbf{\Psi}\left(\mathbf{x}_k + \gamma\mathbf{F}^*\mathbf{U}^T\left(\mathbf{y} - \mathbf{UFx}_k\right)\right)\right)$

Since pFISTA converges much faster than pISTA both theoretically and numerically, we mainly discuss pFISTA in the rest of this paper.

---

**Algorithm 1**: pISTA for tight frames in CS-MRI

Parameters: $\lambda, \gamma$

Initialization: $\mathbf{x}_0$

While not converge, do

$\quad\mathbf{x}_{k+1} = \mathbf{\Psi}^*T_{\gamma\lambda}\left(\mathbf{\Psi}\left(\mathbf{x}_k + \gamma\mathbf{F}^*\mathbf{U}^T\left(\mathbf{y} - \mathbf{UFx}_k\right)\right)\right)$

Output: $\mathbf{x}$

---

**Algorithm 2**: pFISTA for tight frames in CS-MRI

Parameters: $\lambda, \gamma$

Initialization: $t_0 = 1, \mathbf{x}_0, \hat{\mathbf{x}}_0$

While not converge, do

$\quad\mathbf{x}_{k+1} = \mathbf{\Psi}^*T_{\gamma\lambda}\left(\mathbf{\Psi}\left(\hat{\mathbf{x}}_k + \gamma\mathbf{F}^*\mathbf{U}^T\left(\mathbf{y} - \mathbf{UF}\hat{\mathbf{x}}_k\right)\right)\right)$

$\quad t_{k+1} = \dfrac{1 + \sqrt{1 + 4t_k^2}}{2}$

$\quad\hat{\mathbf{x}}_{k+1} = \mathbf{x}_{k+1} + \dfrac{t_k - 1}{t_{k+1}}(\mathbf{x}_{k+1} - \mathbf{x}_k)$

Output: $\mathbf{x}$

---

The pFISTA owns the following advantages:

*1) Low Memory Consumption*

The proposed pFISTA operates on images instead of tight frame coefficients in the original FISTA. Furthermore, pFISTA does not require any auxiliary frame coefficients used in popular analysis model solvers of such as ADMM [29-31] (a.k.a. split Bregman [32] for tight frames). These can significantly reduce the memory consumption since coefficients need more memory than images for a redundant tight frame system. For example, in our numerical experiment, the number of redundant wavelet coefficients for an image is 13 times as many as the number of image pixels. Thus, the pFISTA is memory saving for large scale data and highly redundant systems.

*2) Simplicity*

The simplicity of pFISTA means that besides the regularization parameter, there is only one free parameter, the step size $\gamma$, needs to be set. Besides, it will be shown that this parameter only affects the convergence speed but not change the empirical image reconstruction errors. We recommend users to set $\gamma = 1$ for low reconstruction error and fast convergence speed in tight frames based CS-MRI applications.

*3) Fast Computation and Superior Image Quality*

Fast computation means that pFISTA inherits fast convergence of FISTA as it will be shown in convergence analysis and experiments. Moreover, since pFISTA is an approximate solver for analysis models, it gives images with better quality than FISTA for synthesis models.

*B. Convergence Analysis*

In this section, we will analyze the convergence of both pISTA and pFISTA for tight frames in CS-MRI.

**Theorem 1**: Let $\{\mathbf{x}_k\}$ be generated by pISTA. Providing that the step size $0 < \gamma \leq 1$ and $\mathbf{\Psi}$ is a tight frame, the sequence $\{\mathbf{\alpha}_k\} = \{\mathbf{\Psi x}_k\}$ converges to the solution of

$$\min_{\mathbf{\alpha}} \lambda\|\mathbf{\alpha}\|_1 + \frac{1}{2}\|\mathbf{y} - \mathbf{UF\Psi}^*\mathbf{\alpha}\|_2^2 + \frac{1}{2\gamma}\|(\mathbf{I} - \mathbf{\Psi\Psi}^*)\mathbf{\alpha}\|_2^2 \quad (27)$$

with the speed as

$$F(\mathbf{\alpha}_k) - F(\bar{\mathbf{\alpha}}) \leq \frac{1}{2\gamma k}\|\mathbf{\alpha}_0 - \bar{\mathbf{\alpha}}\|_2^2 \quad (28)$$

where $\bar{\mathbf{\alpha}}$ is a solution of (27) and $F(\cdot)$ is the objective function in (27).

**Theorem 2**: Let $\{\mathbf{x}_k\}$ be generated by pFISTA. Providing that the step size $0 < \gamma \leq 1$ and $\mathbf{\Psi}$ is a tight frame, the sequence $\{\mathbf{\alpha}_k\} = \{\mathbf{\Psi x}_k\}$ converges to the solution of (27) with the speed as

$$F(\mathbf{\alpha}_k) - F(\bar{\mathbf{\alpha}}) \leq \frac{2}{\gamma(k+1)^2}\|\mathbf{\alpha}_0 - \bar{\mathbf{\alpha}}\|_2^2 \quad (29)$$

where $\bar{\mathbf{\alpha}}$ is a solution of (27) and $F(\cdot)$ is the objective function in (27).

Proof of **Theorem 1** and **Theorem 2**:

Let us denote

$$\begin{aligned}h(\mathbf{\alpha}) &= \lambda\|\mathbf{\alpha}\|_1 \\ u(\mathbf{\alpha}) &= \frac{1}{2}\|\mathbf{y} - \mathbf{UF\Psi}^*\mathbf{\alpha}\|_2^2 + \frac{1}{2\gamma}\|(\mathbf{I} - \mathbf{\Psi\Psi}^*)\mathbf{\alpha}\|_2^2\end{aligned} \quad (30)$$

Then applying proximal algorithm [22, 26, 27] to (27) with step size $\gamma$ results in the following iterations

$$\boldsymbol{\alpha}_{k+1} = \text{prox}_{\gamma h}\left(\boldsymbol{\alpha}_k - \gamma \nabla u(\boldsymbol{\alpha}_k)\right)$$
$$= T_{\gamma \lambda}\left(\boldsymbol{\Psi}\left(\boldsymbol{\Psi}^*\boldsymbol{\alpha}_k + \gamma \mathbf{F}^*\mathbf{U}^T\left(\mathbf{y} - \mathbf{UF}\boldsymbol{\Psi}^*\boldsymbol{\alpha}_k\right)\right)\right). \quad (31)$$

Multiplying both sides by $\boldsymbol{\Psi}^*$ and letting $\mathbf{x}_k = \boldsymbol{\Psi}^*\boldsymbol{\alpha}_k$, we get

$$\mathbf{x}_{k+1} = \boldsymbol{\Psi}^* T_{\gamma \lambda}\left(\boldsymbol{\Psi}\left(\mathbf{x}_k + \gamma \mathbf{F}^*\mathbf{U}^T\left(\mathbf{y} - \mathbf{UF}\mathbf{x}_k\right)\right)\right). \quad (32)$$

Note that this is exactly the same iteration (25) as in both pISTA and pFISTA.

The next question is that can we insure the convergence? This question is directly related to the Lipschitz constant of the gradient $\nabla u$ which is defined as

$$L(\gamma) = L(\nabla u) = \left\|\boldsymbol{\Psi}\mathbf{F}^*\mathbf{U}^T\mathbf{UF}\boldsymbol{\Psi}^* + \frac{1}{\gamma}(\mathbf{I} - \boldsymbol{\Psi}\boldsymbol{\Psi}^*)\right\|_2. \quad (33)$$

According to [26], if the step size satisfies

$$\gamma \leq 1/L(\gamma), \quad (34)$$

or equivalently

$$L(\gamma) \leq 1/\gamma, \quad (35)$$

then both pISTA and pFISTA will converge.

We will prove that

$$\begin{cases} L(\gamma) = 1/\gamma, & 0 < \gamma \leq 1 \\ L(\gamma) \leq 1, & \gamma > 1 \end{cases}. \quad (36)$$

Proof of (36):

For simplicity, we denote that

$$\mathbf{B} = \boldsymbol{\Psi}\mathbf{F}^*\mathbf{U}^T\mathbf{UF}\boldsymbol{\Psi}^* - \frac{1}{\gamma}\boldsymbol{\Psi}\boldsymbol{\Psi}^* = \boldsymbol{\Psi}\mathbf{F}^*\left(\mathbf{U}^T\mathbf{U} - \frac{1}{\gamma}\mathbf{I}\right)\mathbf{F}\boldsymbol{\Psi}^*, \quad (37)$$

then we have

$$L(\gamma) = \left\|\mathbf{B} + \frac{1}{\gamma}\mathbf{I}\right\|_2 = \max_i \left\{\left|\lambda_i(\mathbf{B}) + \frac{1}{\gamma}\right|\right\} \quad (38)$$

where $\lambda_i(\mathbf{B})$ means the $i^{\text{th}}$ eigenvalue of $\mathbf{B}$ because $\mathbf{B} + 1/\gamma \mathbf{I}$ is a Hermitian matrix. Therefore, we need to analyze $\lambda_i(\mathbf{B})$. By using the tight frame property, we have

$$\boldsymbol{\Psi}\mathbf{F}^*\left(\mathbf{U}^T\mathbf{U} - \frac{1}{\gamma}\mathbf{I}\right)\mathbf{F}\boldsymbol{\Psi}^*\boldsymbol{\alpha} = \lambda \boldsymbol{\alpha}$$
$$\Rightarrow \left(\mathbf{U}^T\mathbf{U} - \frac{1}{\gamma}\mathbf{I}\right)\mathbf{F}\boldsymbol{\Psi}^*\boldsymbol{\alpha} = \lambda \mathbf{F}\boldsymbol{\Psi}^*\boldsymbol{\alpha}, \quad (39)$$

which indicates that all non-zero eigenvalues of $\mathbf{B}$ satisfy

$$\lambda_i(\mathbf{B}) \in \left\{\lambda\left(\mathbf{U}^T\mathbf{U} - \frac{1}{\gamma}\mathbf{I}\right)\right\} = \left\{-\frac{1}{\gamma}, 1 - \frac{1}{\gamma}\right\}. \quad (40)$$

Due to the redundancy, there exists $\boldsymbol{\alpha} \neq \mathbf{0}$ such that $\boldsymbol{\Psi}^*\boldsymbol{\alpha} = \mathbf{0}$. Thus there are zero eigenvalues of $\mathbf{B}$. Together, we have

$$\begin{aligned} & \lambda_i(\mathbf{B}) = 0 \quad \text{for at least one choice of } i \\ & \& \lambda_i(\mathbf{B}) \in \left\{-\frac{1}{\gamma}, 1 - \frac{1}{\gamma}\right\} \text{ for other choices of } i \end{aligned}. \quad (41)$$

Equation (41) indicates that

$$L(\gamma) = \max_i \left\{\left|\lambda_i(\mathbf{B}) + \frac{1}{\gamma}\right|\right\} = \frac{1}{\gamma}, \quad 0 < \gamma \leq 1$$
$$L(\gamma) = \max_i \left\{\left|\lambda_i(\mathbf{B}) + \frac{1}{\gamma}\right|\right\} \leq 1, \quad \gamma > 1 \quad (42)$$

Done proof of (36).

The relation (36) means that, when $0 < \gamma \leq 1$, one has $L(\gamma) = 1/\gamma$. This together with [26] implies that pISTA and pFISTA will converge with speed described in (28) and (29). Done proof of **Theorem 1** and **Theorem 2**.

*C. Connections with Balanced Sparse Model*

As shown in last section, both pISTA and pFISTA converge to an approximate model (27) instead of the exact analysis model (9) or (14). The model (27) is not new in general image restoration and it was called the balanced sparse model that balances solutions between synthesis and analysis sparse models [33-35]. The performance of balanced sparse models in CS-MRI was studied by Liu *et al.* [10] and their results show that the reconstruced errors and images of balanced sparse models are comparable to those of analysis models for all the tested tight frames in CS-MRI. Shen *et al.* [33] proposed an accelerated proximal gradient algorithm (APG) to solve balanced sparse models in common image restoration tasks, including deblurring, denoising and component decomposition, but not CS-MRI problems. Although from different perspectives, it turns out that pFISTA coincides with APG when the linear operator is chosen as undersampling Fourier operator. However, pFISTA is not a trivial extension for following reasons:

1) Although tight frames are shown to improve the image quality significantly in CS-MRI, but how to solve tight frames-based MRI image reconstruction fast and with minimal free parameters is still unknown. The proposed pFISTA only introduces one parameter, the step size, and experiments show that reconstruction errors are insensitive to this parameter (See Section III).

2) The APG algorithm is formulated and implemented in frame coefficients domain, and it needs to store copies of all redundant tight frame coefficients. Our pFISTA works in image domain, and there is no need to store any tight frame coefficients. Therefore, the pFISTA can significantly reduce memory consumption for highly redundant systems.

These two properties allow users in MRI to easily set algorithm parameters and utilize different tight frames for high quality image reconstruction from undersampled k-space data.

III. NUMERICAL EXPERIMENTS

We will verify the effectiveness and efficiency of the proposed pFISTA for tight frames in CS-MRI by comparing it with FISTA (for synthesis model) [26] and the state-of-art SFISTA (for approximated analysis model) [28].

We will conduct experiments on three typical MRI data: a water phantom image, a T2-weighted brain image and a T1-weighted brain image as shown in Fig. 2. The water phantom image was acquired at 7T Varian MRI system (Varian,

Palo Alto, CA, USA) with single coil using the spin echo sequence (TR/TE = 2000/100 ms, FOV 80×80 mm$^2$, and slice thickness = 2 mm). The T2-weighted brain image was acquired from a healthy volunteer at a 3T Siemens Trio Tim MRI scanner with 32 coils using the T2-weighted turbo spin echo sequence (TR/TE = 6100/99 ms, FOV = 220×220 mm$^2$, slice thickness = 3 mm). The T1-weighted brain image was acquired from a healthy volunteer at a 1.5T Philips MRI scanner with 8 coils using sequence parameters (TR/TE = 1700/390 ms, FOV = 230×230 mm$^2$, slice thickness = 5 mm). For the multi-channel brain imaging data, we first did the SENSE reconstruction with reduction factor 1 to compose the ground truth images to emulate the single-channel MRI data [36, 37]. These ground truth images are with real and imaginary parts and we do the reconstruction from the retrospectively undersampled Fourier coefficients of these images. All three images are of size 256×256 and with real and imaginary parts. The k-space undersampling is simulated by using the mask in Fig. 2 (d) with 40% of k-space data being sampled. Note that in our application on 2D imaging, instead of the fully 2D randomly sampling in [4,6], the undersampling here is only along the phase encoding dimension because the frequency encoding dimension is not time-consuming and is unworthy of undersampling [1,5,38]. The i.i.d. complex Gaussian noise with standard deviation $\sigma = 0.01$ is added to the k-space of all imaging data to test the robustness of the algorithms to noise which is commonly encountered in MRI [39, 40].

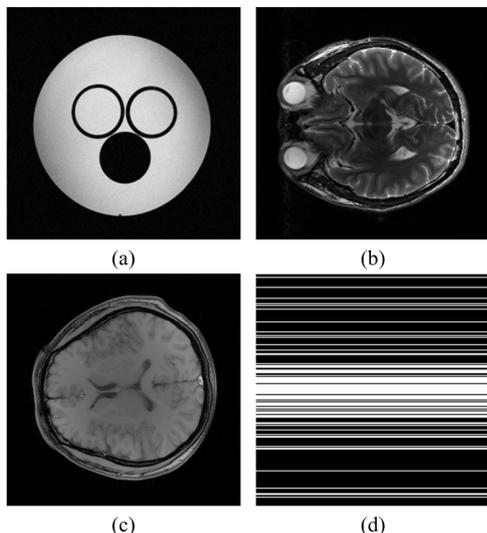

Fig. 2. Experimental datasets. (a) is a water phantom, (b) is a T2-weighted brain image, (c) is a T1-weighted brain image, (d) is the sampling mask with 40% data points are sampled.

All numerical experiments are conducted on a Dell PC running Windows 7 operating system with Intel Core i7 2600 CPU. For quantitative comparison, we adopt the relative $l_2$ norm error (RLNE) defined as

$$\text{RLNE} := \|\hat{\mathbf{x}} - \mathbf{x}\|_2 / \|\mathbf{x}\|_2, \qquad (43)$$

where $\mathbf{x}$ is the ground truth image and $\hat{\mathbf{x}}$ is the reconstructed image. This criteria was previously used in MR image reconstructions [5, 36].

There are several parameters to tune in the algorithms, namely the step size $\gamma_F$ for FISTA, step size $\gamma_S$ and smooth approximate parameter $\mu$ for SFISTA, and step size $\gamma_p$ for pFISTA. We set $\gamma_F = 1$ and $\gamma_S = 1/(1+1/\mu)$ for fast convergence [26, 28] and $\mu$ is adjusted according to [28]. We choose $\gamma_p = 1$ for both promising reconstruction performance and fast speed. The regularization parameters, $\lambda_F$ of FISTA, $\lambda_S$ of SFISTA and $\lambda_p$ of pFISTA, are manually tuned so that we have an optimal reconstruction error in terms of RLNE.

### A. Main Results

The shift-invariant discrete wavelet transform (SIDWT) implemented in Rice Wavelet Toolbox [41] is adopted as a typical tight frame in simulation. SIDWT is also known as undecimated/translation-invariant/fully redundant wavelets. Without shift-invariance property, the signal restored using the orthogonal discrete wavelet transform will exhibit much more artifacts in denoising [11, 12]. Within the field of MRI, some researchers have utilized SIDWT to reconstruct MR images and found it superior than its orthogonal counterpart in noise suppression and artifacts reduction [5, 36, 42-44]. In all the experiments, Daubechies wavelets with 4 decomposition levels are utilized in SIDWT. The regularization parameters are tuned as $\lambda_F = \lambda_S = \lambda_p = 0.01$, 0.001 and 0.001 for the water phantom, the T2-weighted brain image and the T1-weighted brain image, respectively.

As for reconstructed images, the phenomena of the experiments on the three different MRI datasets are consistent. As shown in Figs. 3-5, the reconstructed images of FISTA exhibit obvious artifacts which are suppressed much better by using SFISTA and pFISTA. The FISTA reaches to a higher reconstruction error while the latter two produce close errors. The original FISTA solves the synthesis sparse models, which usually produced sub-optimal results compared with analysis model, solved by SFISTA, and balanced models, solved by pFISTA. This observation is consistent with other researchers [8, 14, 17, 18, 34]. Note that we set $\mu = 10^{-4}/\lambda_S$ here to obtain the best reconstructed images of SFISTA according to [28].

The main difference between SFISTA and pFISTA is the convergence. In order to fairly compare these two algorithms, we show the convergence of SFISTA against $\mu$. Fig. 6 shows that pFISTA converges faster than SFISTA with $\mu = 1$ while achieving comparable errors.

### B. Discussions

*1) Sensitivity of pFISTA to the Step Size*

As shown in Section II(B), the step size $\gamma_p$ is not only a parameter that affects the convergence speed but also determines the model (27) that pFISTA converges to. In this section, we numerically investigate how the step size $\gamma_p$ affects the convergence and reconstruction.

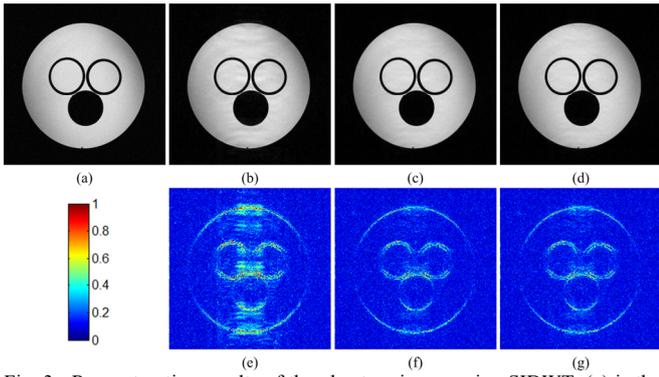

Fig. 3. Reconstruction results of the phantom image using SIDWT. (a) is the ground truth image. (b)-(d) are reconstructed images of FISTA, SFISTA and pFISTA with the RLNE errors are 0.070, 0.054 and 0.055, respectively. (e)-(g) are 5x scaled difference images of (b)-(d) to (a).

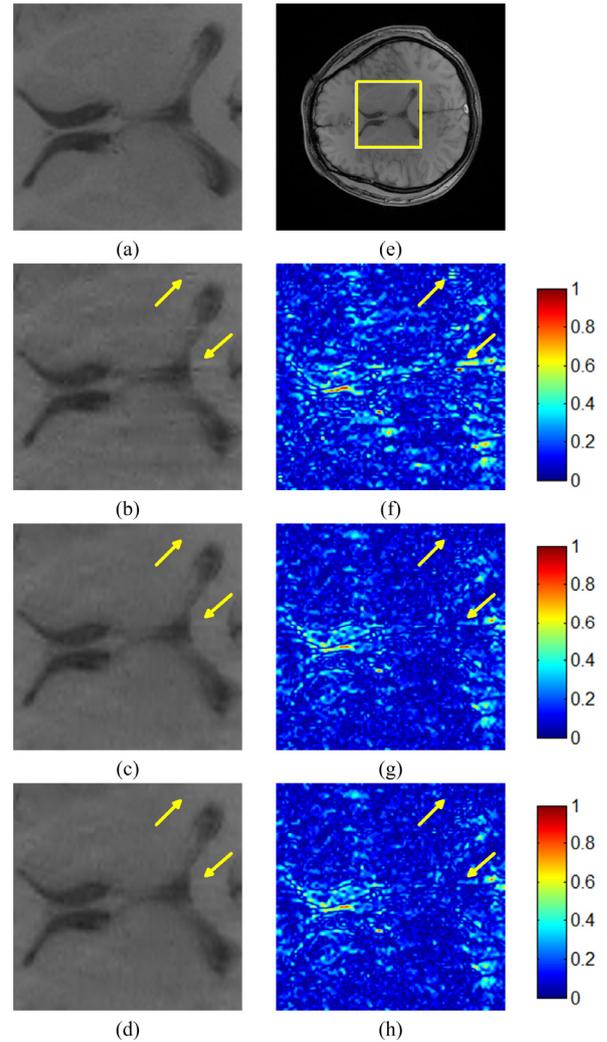

Fig. 5. Reconstruction results of the T1-weighted brain image using SIDWT. (a) is the zoom-in region marked in the ground truth image in (e). (b)-(d) are zoom-in regions of reconstructed images of FISTA, SFISTA and pFISTA with the RLNE errors are 0.098, 0.086 and 0.086, respectively. (f)-(h) are 10x scaled difference images of (b)-(d) to (a).

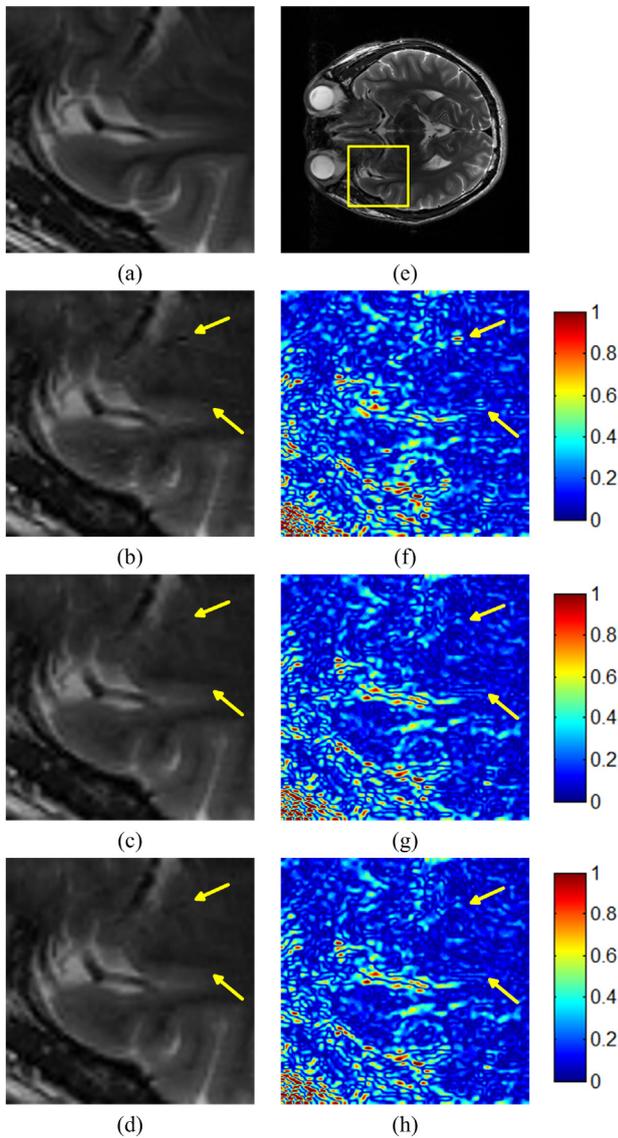

Fig. 4. Reconstruction results of the T2-weighted brain image using SIDWT. (a) is the zoom-in region marked in the ground truth image in (e). (b)-(d) are zoom-in regions of reconstructed images of FISTA, SFISTA and pFISTA with the RLNE errors are 0.138, 0.127 and 0.127, respectively. (f)-(h) are 10x scaled difference images of (b)-(d) to (a).

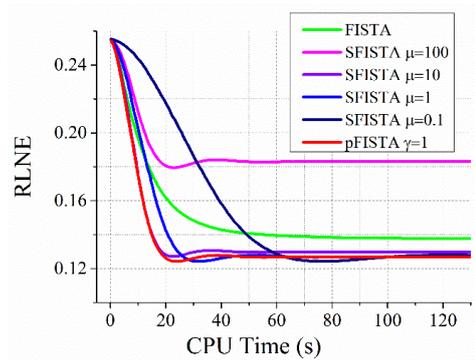

Fig. 6. Comparison of empirical convergence of FISTA, SFISTA and pFISTA using SIDWT.

Fig. 7 shows that with a larger $\gamma_p$, the convergence speed is faster while the final RLNE is almost not changed. Therefore, the reconstruction error of pFISTA is insensitive to the change of step size. This is why we recommend setting the $\gamma_p = 1$ for both promising reconstruction performance and fast speed in tight frames-based CS-MRI.

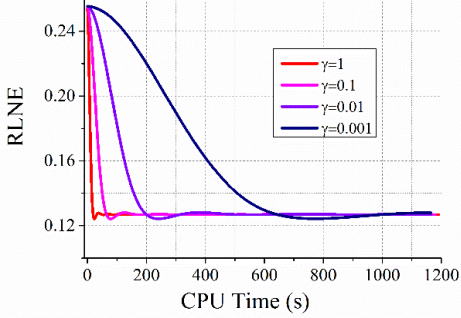

Fig. 7. Empirical convergence of pFISTA with different step sizes γ using SIDWT.

*2) Experiments on Other Tight Frames*

Different tight frames may affect convergence speed and reconstruction errors. Thus, it is worthy to discuss some other tight frames. In this section, we conduct experiments on the data in Fig. 2 (b) based on other three tight frames: contourlet [4, 45], patch based directional wavelets (PBDW) [5] and framelet [46-48]. Both contourlet and PBDW explore the geometric information to further sparsify MR images thus are good at preserving the edge of MR images [4, 5, 45]. The filters of contourlet, PBDW and framelet are ladder structure filters, Haar wavelets and Haar wavelets, respectively. The decomposition levels of contourlet, PBDW and framelet are [5,4,4,3], 3 and 1, respectively. The regularization parameters for contourlet, PBDW and framelet are $\lambda_F = \lambda_S = \lambda_p = 0.01$, $\lambda_F = \lambda_S = \lambda_p = 0.001$ and $\lambda_F = \lambda_S = \lambda_p = 0.001$, respectively. To allow reproducing the results, the code will be released at http://www.quxiaobo.org/project/pFISTA_MRI/Demo_pFISTA_MRI.zip .

Fig. 8 shows that, when using contourlet, PBDW and framelet as sparsifying transforms in CS-MRI, we observe that within each tight-frame: 1) RLNE errors of SFISTA and pFISTA are smaller than FISTA; 2) pFISTA converges faster than SFISTA when they achieve comparable RLNE errors; 3) the reconstruction error of pFISTA is stable to the step size. Thus, the advantages of pFISTA over SFISTA and FISTA do not depend on the choice of tight frames.

*3) Comparison with the Exact Analysis Model*

The convergence analysis in the Section II(B) proves that pFISTA does not solve the exact analysis model in (9). Instead, it solves the approximate analysis model in (27). In our previous work [14], we studied the performance of the model (27) in CS-MRI and found that reconstructions of (27) are comparable to those of exact analysis models in numerical experiments [14]. In this section, we conduct experiments to compare the reconstructions of pFISTA to that of the exact analysis model in (9) which is solved by alternating direction

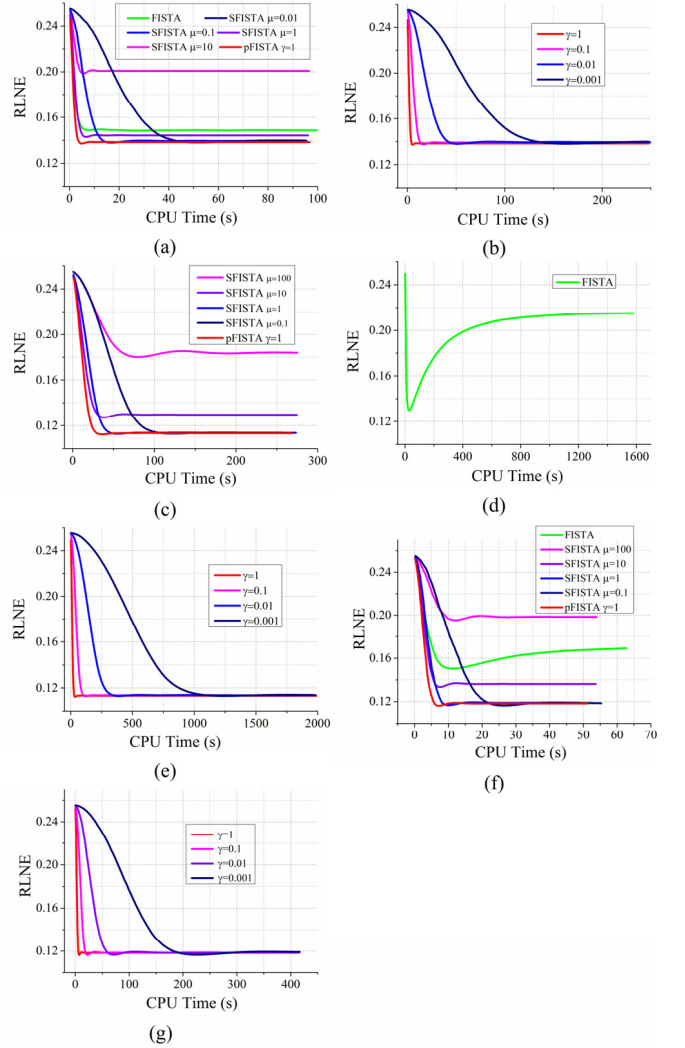

Fig. 8. Empirical convergence using other tight frames: contourlet, PBDW and framelet. (a) is RLNEs of FISTA, SFISTA and pFISTA using contourlet, (b) is RLNEs of pFISTA with different γ using coutourlet, (c) is RLNEs of SFISTA and pFISTA using PBDW, (d) is RLNEs of FISTA using PBDW, (e) is RLNEs of pFISTA with different γ using PBDW, (f) is RLNEs of FISTA, SFISTA and pFISTA using framelet, (g) is RLNEs of pFISTA with different γ using framelet.

method of multipliers (ADMM) [49, 50].

Fig. 9 shows the reconstructed errors of ADMM (exact analysis), SFISTA (approximated analysis) and pFISTA (the proposed) with SIDWT. The reconstruction errors of both SFISTA and pFISTA are almost the same as that of the exact analysis model solved using ADMM. Note that we plotted the convergence curves of ADMM with multiple penalty parameters $\rho$, since the convergence speed of ADMM is sensitive to this parameter [49-51]. With $\rho = 0.01$, ADMM converges fastest but a larger or smaller converges much slower in this case. To the best of our knowledge, it is still unknown to tune an optimal $\rho$ of ADMM in CS-MRI [49-51].

We also conduct experiments to compare the performance using other tight frames and MRI data used before. Results listed in Table I imply that the approximate analysis model solved by pFISTA and the approximate analysis model solved by SFISTA, achieve almost the same reconstruction errors as exact analysis model solved by ADMM, when the undersampled data and sparsifying transforms are fixed.

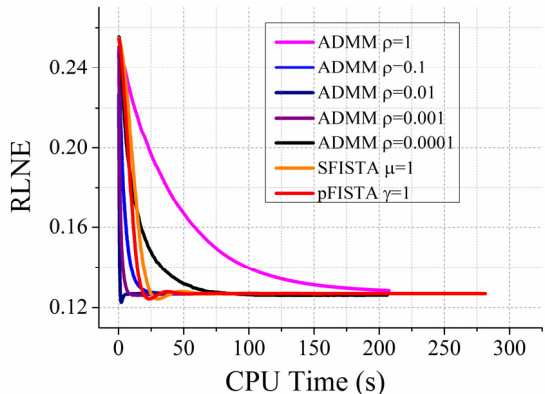

Fig. 9. Reconstructed errors of ADMM, SFISTA and pFISTA. Note: The T2-weighted image in Fig. 2 (b), undersampling pattern in Fig. 2 (d) are adopted in the experiment.

TABLE I
RECONSTRUCTIED ERRORS, RLNES, USING ADMM/SFISTA/PFISTA.

|  | Phantom | T1-weighted | T2-weighted |
|---|---|---|---|
| SIDWT | 0.054/0.054/0.054 | 0.084/0.085/0.084 | 0.126/0.126/0.126 |
| Contourlet | 0.068/0.068/0.068 | 0.091/0.091/0.090 | 0.136/0.135/0.135 |
| Framelet | 0.051/0.051/0.051 | 0.083/0.084/0.084 | 0.118/0.119/0.118 |
| PBDW | 0.062/0.062/0.062 | 0.077/0.077/0.077 | 0.114/0.113/0.114 |

Note: Fig. 2(d) is adopted as the undersampling pattern.

*4) Experiments on Other Undersampling Patterns*

Undersampling patterns are very important to reduce reconstruction errors in CS-MRI. In aforementioned experiments, the sampling pattern in Fig. 2 (d) is a 1D undersampling only along the phase encoding direction but not the frequency encoding direction. This is because, in common MRI experiments, frequency encoding direction is not very time-consuming and it is not worthy undersampling this dimension. Anyway, our algorithm also adapts to other undersampling patterns. In this section, we conduct experiments with the 2D undersampling pattern (left of Fig. 10), which emulates the 2D phase encodings in 3D imaging, and the radial undersampling pattern (right of Fig. 10).

Results in Table II imply that, when using 2D undersampling pattern with the same amount of 40% data sampled in Fig. 2(d), reconstruction errors are significantly reduced, namely from 0.126 to 0.099 for SIDWT and from 0.113 to 0.086 for PBDW on the T2-weighted image. For the radial sampling with only 30% data, the reconstructed errors are lower than that for 1D undersampling with 40% data.

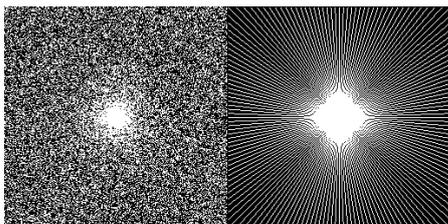

Fig. 10. Other undersampling patterns. Left is 2D undersampling pattern with 40% data being sampled. Right is radial undersampling pattern with 30% data being sampled.

TABLE II
RECONSTRUCTION ERRORS WITH FULLY 2D / RADIAL UNDERSAMPLING.

|  | Phantom | T1-weighted | T2-weighted |
|---|---|---|---|
| SIDWT | 0.049/0.052 | 0.069/0.078 | 0.099/0.114 |
| Contourlet | 0.055/0.053 | 0.078/0.083 | 0.120/0.127 |
| Framelet | 0.048/0.051 | 0.068/0.073 | 0.089/0.104 |
| PBDW | 0.060/0.063 | 0.063/0.070 | 0.086/0.102 |

Note: The SIDWT is used as sparsifying transform and the undersampling patterns are shown in Fig. 10.

IV. CONCLUSION

We propose a projected iterative soft-threshoding algorithm (pISTA) and futher accelerate it with the same strategy as FISTA, namely pFISTA, to solve sparse image reconstruction from undersampled measurements in fast magnetic resonance imaging. We theoretically prove that the proposed algorithm converges to the balanced sparse model. Numerical results show that pFISTA achieves better reconstruction than FISTA for synthesis sparse model and converges faster or comparable to the state-of-art SFISTA for analysis sparse model. One main advantage of pFISTA is that reconstructed errors are stable to the step size, thus allowing widely usage for different tight frames in magnetic resonance image reconstructions. In te future, the convergence of pFISTA for general frames/dictionaries will be analyzeed and this algorithm will be used for other advanced adaptively sparse representations [52-53] in compressed sensing MRI.

Appendix

Proof of equivalence between (14) and (9) in the manuscript.

Denoting that $G(\mathbf{x}) = \lambda \|\mathbf{\Psi x}\|_1 + 1/2 \|\mathbf{y} - \mathbf{UFx}\|_2^2$, then one has

$$\begin{aligned}
&\min_{\boldsymbol{\alpha} \in \text{Range}(\boldsymbol{\Psi})} \lambda \|\boldsymbol{\alpha}\|_1 + \frac{1}{2} \|\mathbf{y} - \mathbf{UF\Phi\alpha}\|_2^2 \\
&\stackrel{(a)}{=} \min_{\boldsymbol{\alpha} \in \text{Range}(\boldsymbol{\Psi})} \lambda \|\boldsymbol{\Psi\Phi\alpha}\|_1 + \frac{1}{2} \|\mathbf{y} - \mathbf{UF\Phi\alpha}\|_2^2 \\
&\stackrel{(b)}{=} \min_{\boldsymbol{\alpha} \in \text{Range}(\boldsymbol{\Psi})} G(\boldsymbol{\Phi\alpha}) \\
&\stackrel{(c)}{=} \min_{\mathbf{x} \in \Omega} G(\mathbf{x})
\end{aligned} \quad (45)$$

with $\Omega = \{\boldsymbol{\Phi\alpha} \mid \boldsymbol{\alpha} \in \text{Range}(\boldsymbol{\Psi})\}$ where (a) from the property (7) for $\boldsymbol{\alpha} \in \text{Range}(\boldsymbol{\Psi})$, (b) and (c) are straightforward based on the definition of $G(\cdot)$ and $\Omega$. Next, we show that $\Omega = \mathbb{C}^N$. On one hand, we have

$$\mathbf{x} \in \mathbb{C}^N \stackrel{\boldsymbol{\Phi\Psi x} = \mathbf{x}}{\Rightarrow} \mathbf{x} \in \Omega \text{ with } \boldsymbol{\alpha} = \boldsymbol{\Psi x}. \quad (46)$$

On the other hand, we have

$$\begin{aligned}
&\mathbf{x} \in \Omega \\
&\Rightarrow \mathbf{x} = \boldsymbol{\Phi\alpha} \text{ for some } \boldsymbol{\alpha} \in \text{Range}(\boldsymbol{\Psi}) \\
&\Rightarrow \mathbf{x} = \boldsymbol{\Phi\alpha} \text{ with } \boldsymbol{\alpha} = \boldsymbol{\Psi\tilde{x}} \text{ for some } \tilde{\mathbf{x}} \in \mathbb{C}^N \\
&\Rightarrow \mathbf{x} = \boldsymbol{\Phi\Psi\tilde{x}} = \tilde{\mathbf{x}} \text{ for some } \tilde{\mathbf{x}} \in \mathbb{C}^N \\
&\Rightarrow \mathbf{x} \in \mathbb{C}^N
\end{aligned} \quad (47)$$

(46) and (47) together leads to $\Omega = \mathbb{C}^N$. This together with (45) leads to

$$\begin{aligned}
&\min_{\boldsymbol{\alpha} \in \text{Range}(\boldsymbol{\Psi})} \lambda \|\boldsymbol{\alpha}\|_1 + \frac{1}{2} \|\mathbf{y} - \mathbf{UF\Phi\alpha}\|_2^2 \\
&= \min_{\mathbf{x}} \lambda \|\boldsymbol{\Psi x}\|_1 + \frac{1}{2} \|\mathbf{y} - \mathbf{UFx}\|_2^2
\end{aligned}. \quad (48)$$

If $\boldsymbol{\alpha}^*$ is a solution of (14) and $\mathbf{x}^*$ is a solution of (9), then one has

$$G(\boldsymbol{\Phi\alpha}^*) \stackrel{(d)}{=} G(\mathbf{x}^*) \stackrel{(e)}{=} G(\boldsymbol{\Phi\Psi x}^*) \quad (49)$$

where (d) from the second equation in (45) and (48), (e) from (5). Therefore, $\boldsymbol{\Phi\alpha}^*$ is also a solution of the analysis model (9) and $\boldsymbol{\Psi x}^*$ is also a solution of the synthesis-like model (14). Done proof.